# Improvement of critical current density in Fe-sheathed $MgB_2$ tapes by $ZrSi_2$, $ZrB_2$ and $WSi_2$ doping

Yanwei Ma*, H. Kumakura, A. Matsumoto, H. Hatakeyama, K. Togano

Superconducting Materials Center, National institute for Materials Science, 1-2-1 Sengen, Tsukuba, Ibaraki 305-0047, Japan

* E-mail: mayanwei@yahoo.com

**Abstract**

Fe-sheathed $MgB_2$ tapes were prepared through the *in situ* powder-in-tube technique by 5 at.% $ZrSi_2$, $ZrB_2$ and $WSi_2$ doping, respectively. The doping effect of these compounds on the microstructure and superconducting properties of $MgB_2$ tapes has been investigated by using x-ray diffraction, scanning electron microscope, transport measurements and DC susceptibility measurements. Compared to the undoped samples, Jc for all the doped samples were much improved; the best result in terms of Jc was achieved for $ZrSi_2$ doping, by up to a factor of 3.4 at 4.2 K in magnetic fields up to 12 T. Moreover, these dopants did not significantly decrease the transition temperature. The Jc-B curves of $WSi_2$-doped tapes show better performance in higher magnetic fields in comparison to undoped tapes, suggesting that pinning centers effective in a high-field region were possibly introduced.

**Introduction**

The discovery of superconductivity in $MgB_2$ with a transition temperature (Tc) of 39 K [1] has led to great progress researches in applied superconductivity because of low materials cost. The crystal structure of this material is an hexagonal $AlB_2$ type structure consisting of alternating layers of Mg atoms and boron honeycomb layers. Unlike the case of copper oxide superconductors $MgB_2$ has no weak-link problem at grain boundaries [2], and is thus a promising candidate for engineering applications in the temperature range of liquid hydrogen (20-30K) where the conventional superconductors cannot play any role due to low Tc. To achieve high critical current density, different techniques have been developed. Among these techniques, the powder-in-tube (PIT) method has been demonstrated, either with [3-5] or without [6-7] recrystallization after deformation. Iron and its alloy have been found to be useful not only due to non-poisoning to $MgB_2$ but also due to its ductility, low cost and light weight. Extensive research efforts have been made to improve the Jc of Fe-sheathed $MgB_2$ tapes [3, 4]. High critical current density Jc up to $10^4 – 10^5$ A/cm$^2$ at 10 K was obtained. However, the Jc value of $MgB_2$ tapes is relatively low compared to the conventional low temperature superconductors, due to poor grain connections and the lack of flux pinning centers in this material.

On the other hand, the chemical doping is found to be easily controlled and highly efficient in improving Jc and flux pinning in high-Tc superconductors [19-20]. Very recently, Dou [8] and Driscoll et al. [9] reported that chemical doping with nano-particles (SiC or $Y_2O_3$) into $MgB_2$ could significantly enhance Jc. Cimberle et al. [10] found that Jc increased somewhat for Li-, Al-, and Si-doped samples. Likewise, Feng claimed much higher Jc for Ti- and Zr-doped samples at low fields [11-12].

In this work, we have fabricated separate $ZrSi_2$-, $ZrB_2$- and $WSi_2$-doped $MgB_2$ tapes and investigated those doping effect on the microstructure and Jc of $MgB_2$. Excellent performance of Jc in magnetic fields of up to 12 T has been achieved in doped $MgB_2$ tapes processed at ambient pressure.

**Experimental**

The $MgB_2$ composite tapes with $ZrSi_2$ (or $ZrB_2$ or $WSi_2$) doping were prepared by the *in situ* PIT method with Fe sheath. The doping ratio of Mg:X:B was 5 at.% with X = $ZrSi_2$, $ZrB_2$ and $WSi_2$. Mg (325 mesh, 99.8% in purity), $ZrSi_2$ (or $ZrB_2$ or $WSi_2$)(2~5 µm), and B amorphous (325 mesh, 99.99%) powders were well mixed and ground in air for 1 h. The pure Fe tubes had an outside diameter of 6 mm, a wall thickness of 1.25 mm, and were 5 cm long. One end of the tube was sealed; then the powder mixture was filled into the Fe tube in air. After packing, the other end was crumpled, and this tube was subsequently cold-rolled into a rectangular rod using a groove-rolling machine.

Then, the rods were cold-rolled into a tape with a thickness of about 0.5 mm and a width of about 4 mm. These tapes were cut into short pieces and then heated up to 600° C in 40 min in a flow of Ar. After 1 h heat treatment, the tapes were furnace-cooled to room temperature. Undoped $MgB_2$ tapes were similarly prepared for comparison.

Phase identification was performed by an x-ray diffraction (XRD) method after mechanically peeling off the sheath materials. Microstructural observation was carried out by scanning electron microscopy (SEM). DC magnetization measurements were performed with a superconducting quantum interference device magnetometer (SQUID). Magnetization curves were measured with a vibrating sample magnetometer (VSM). Using a conventional four-probe resistive method, the transport critical current (Ic) of short tapes was measured at 4.2 K. The criterion for the Ic definition was 1 µV/cm. A magnetic field up to 12 T was applied parallel to the tape surface. The Ic measurement was performed for several tape samples to check reproducibility.

**Results and discussion**

The phases present after the heat treatment were determined by XRD with Cu Kα radiation at room temperature, as shown in Fig. 1. The XRD pattern for the undoped samples reveals that $MgB_2$ was obtained as the nearly single phase. Note that the peaks of Fe were contributed from the Fe sheath. The XRD spectrums for $ZrB_2$ and $WSi_2$ samples show that in addition to $MgB_2$, large quantities of pure dopants, $ZrB_2$ and $WSi_2$, are presented. Clearly, $MgB_2$ is inert with respect to $ZrB_2$ and $WSi_2$ at 600 °C. Note that due to the high x-ray scattering factor of the heavy element, Zr and W, the relative intensities of $MgB_2$ peaks are smaller than those of $ZrB_2$ and $WSi_2$ peaks even for the doping level = 5 at.%. By contrast, $ZrSi_2$ samples consist of $MgB_2$ as the main phase. The addition of $ZrSi_2$ leads to the formation of $Zr_3Si_2$ and $Mg_2Si$ as the major impurity phases; there are no peaks corresponding to pure $ZrSi_2$, suggesting that there were reaction between $MgB_2$ and $ZrSi_2$. It is worth noting that the position of $MgB_2$ peaks could not be changed with these three materials doping, which is much different from the results in Zn and Al-doped samples [13-14].

Figure 2 shows the superconducting transition curves for the doped and undoped samples determined by susceptibility measurements. Samples were zero-field cooled and then warmed from 5 K in an applied field of 10 Oe. All the doped tapes show a relatively sharp transitions, suggesting a fairly strong coupling of grains. The highest transition temperature (Tc onset =35.7 K) is observed in the pure $MgB_2$ samples. As can be seen, all doping slightly decreases Tc (less than 1 K), indicating that the dopant incorporates into the MgB$_2$ structure. Superconducting transitions, with an onset at 35.2 K, are seen for the ZrSi$_2$ and ZrB$_2$ samples, while WSi$_2$ samples have Tc onset of 35.0 K. Our observations are in good agreement with previous reports, in which the transition temperature decreases at various rates for different doping with Al, Zr, Ti, Si, Li and SiC [10, 11, 13, 15, 18].

Figure 3 summarizes the Jc values at 4.2 K as a function of magnetic fields for our MgB$_2$/Fe tapes with and without doping. The magnetic field was applied parallel to the tape surface. It can be seen that the Jc values are much improved by doping in MgB$_2$ tapes with ZrSi$_2$, ZrB$_2$ and WSi$_2$, respectively. Clearly, the largest Jc values were achieved in the ZrSi$_2$-doped samples, more than three times higher than the undoped ones. Similarly, in the ZrB$_2$- and WSi$_2$-doped tapes, we also obtained the Jc values being at least twice as large as those of pure tapes. At 4.2 K and in a field of 10 T, the undoped samples showed Jc values of about 820 A/cm$^2$, while the doped tapes showed higher values in the same field, namely, ZrSi$_2$-doped tapes = ~3000 A/cm$^2$; WSi$_2$ tapes: ~1900 A/cm$^2$; ZrB$_2$ tapes: ~1700 A/cm$^2$. More importantly, for the WSi$_2$-doped tapes the field dependence of Jc was evidently different for other three samples (undoped, ZrSi$_2$- and ZrB$_2$-doped ). The Jc-B curves became flatter than those of other ones in higher field region. Such a difference can be explained in term of higher flux pinning effect. This is demonstrated by Fig. 4, which shows the field dependence of volume pinning forces Fp(H) at 22 K for the undoped and WSi$_2$-doped tapes. Here Fp(H) is obtained from the hysteresis in magnetization curves and is normalized by the maximum volume pinning force Fp$^{max}$ at the same temperature. Although the positions of the maximum pinning force for both tapes (undoped and WSi$_2$-) are the same, the pinning force over B$_{Fp-max}$ is apparently larger in the WSi$_2$-doped tapes, suggesting that pinning centers effective in a high-field region were possibly introduced.

We investigated the reason for the Jc difference for tapes with and without doping. Figure 5 shows the typical SEM images of the fractured core layers for different doped tapes, as well as the undoped tapes. A rough and porous microstructure is observed in the pure $MgB_2$ tapes. However, all three doped tapes have a higher density with few voids, suggesting the improved coupling of grains. This important enhancement of the $MgB_2$ core quality is believed to be responsible for the higher transport Jc in the doped tapes. The grain boundaries may act as pinning centers in $MgB_2$ as in $Nb_3Sn$ [17]. However, in contrast to previous work on doping for reducing grain size [11, 12], our high magnification images for all tapes indicated that the grain size is almost the same (~ 0.2 µm), which further confirmed that the Jc difference in these tapes is not due to the grain-size difference but due to the improved grain coupling.

The high performance of Jc in the whole range of magnetic fields up to 12 T in the three kinds of doped samples might be largely attributed to densification effect and a good connection between grains. Our results are in consistent with recent reports, in which doping $MgB_2$ with Ti, Si and Zr showed an improvement of Jc, likely mainly due to the densification effect [10-12]. Feng and co-workers claimed that in Zr- and Ti-doped cases, with increasing doping level, the samples become denser, and the Jc increases. However, the improvement in the microstructure is not significant until the doping level reaches 10%. As the doping level increases to 10%, the density of the sample sharply increases and, at the same time, the Jc enhanced significantly. As the doping level further increases to 40%, the Jc decreases rapidly although the samples are still quite dense [11-12]. Based on our present preliminary results, it is speculated that further improvement in Jc will be achieved upon optimization of doping level.

On the other hand, in our present study, the size for added powders is around 2-5 µm, which may be too large for strong flux pinning. This is the reason why the difference of pinning forces between $WSi_2$ doped and undoped tapes is small as shown in Fig. 4. It is evident that incorporating SiC (or $Y_2O_3$) nanoparticles together with Mg and B powders results in the formation of $MgB_2$ with a uniform dispersion of nano-precipitates, which can act as strong pinning centers [8-9]. Thus we can expect that Jc(H) and irreversibility field ($H_{irr}$) can be improved significantly by doping $MgB_2$ with nano-particles $ZrSi_2$, $ZrB_2$ and $WSi_2$, respectively.

In a word, the present results clearly showed that the $ZrSi_2$, $ZrB_2$ and $WSi_2$ doping in $MgB_2$ seems an effective and easily controlled method to improve Jc. This method is very suitable for the industrial scale fabrication of $MgB_2$ tapes and bulks because these doped samples are prepared at ambient pressure.

**Conclusions**

In summary, the Fe-sheathed tapes of the $MgB_2$ pure and doped with $ZrSi_2$, $ZrB_2$ and $WSi_2$ have been prepared through the *in situ* PIT method. The transport Jc measured at low temperature indicates an improvement of the critical current density for all the doped tapes, the best result being achieved by the $ZrSi_2$ doping (about a factor 3.4 in the Jc values). At 4.2 K and in a field of 8 T, the Jc of $ZrSi_2$-doped tapes reaches a higher value about $10^4$ A/cm$^2$. The Jc-B curves of $WSi_2$-doped tapes show better performance in higher magnetic fields in comparison to undoped tapes, suggesting that pinning centers effective in a high-field region were possibly introduced. Densification of $MgB_2$ core was observed for all the doped tapes. This densification is also effective to improve Jc.

**ACKNOWLEDGMENT**

The authors greatly thank Dr. H. Takeya, S. Yu, M. Sumita, T. Nakane, T. Mochiku, H. Kitaguchi for their kind help during measurements.

**Figure captions**

Figure 1 X-ray diffraction patterns for the undoped and all the doped samples heated at 600 °C for 1h. the data were obtained after peeling off the Fe-sheath. The XRD peaks of $MgB_2$ are indexed, and the peaks of $WSi_2$, $ZrB_2$, $Zr_3Si_2$ and $Mg_2Si$ are marked by squares, asterisks, solid circles and open circles, respectively. The peaks of Fe were contributed from the Fe sheath.

Figure 2 Normalized magnetic susceptibility vs temperature for all the doped and undoped tapes. The inset shows the enlarged view near the superconducting transitions.

Figure 3 Jc-B properties of Fe-sheathed undoped and all the doped tapes heated at 600 °C for 1h. The measurements were performed in magnetic fields parallel to the tape surface.

Figure 4 The normalized volume pinning forces of undoped and $WSi_2$-doped taped at 22 K.

Figure 5 SEM images of the fractured $MgB_2$ core layers of Fe-sheathed undoped and all the doped tapes heated at 600 °C for 1h. (a) undoped, (b) $ZrSi_2$, (c) $ZrB_2$, (d)$WSi_2$.

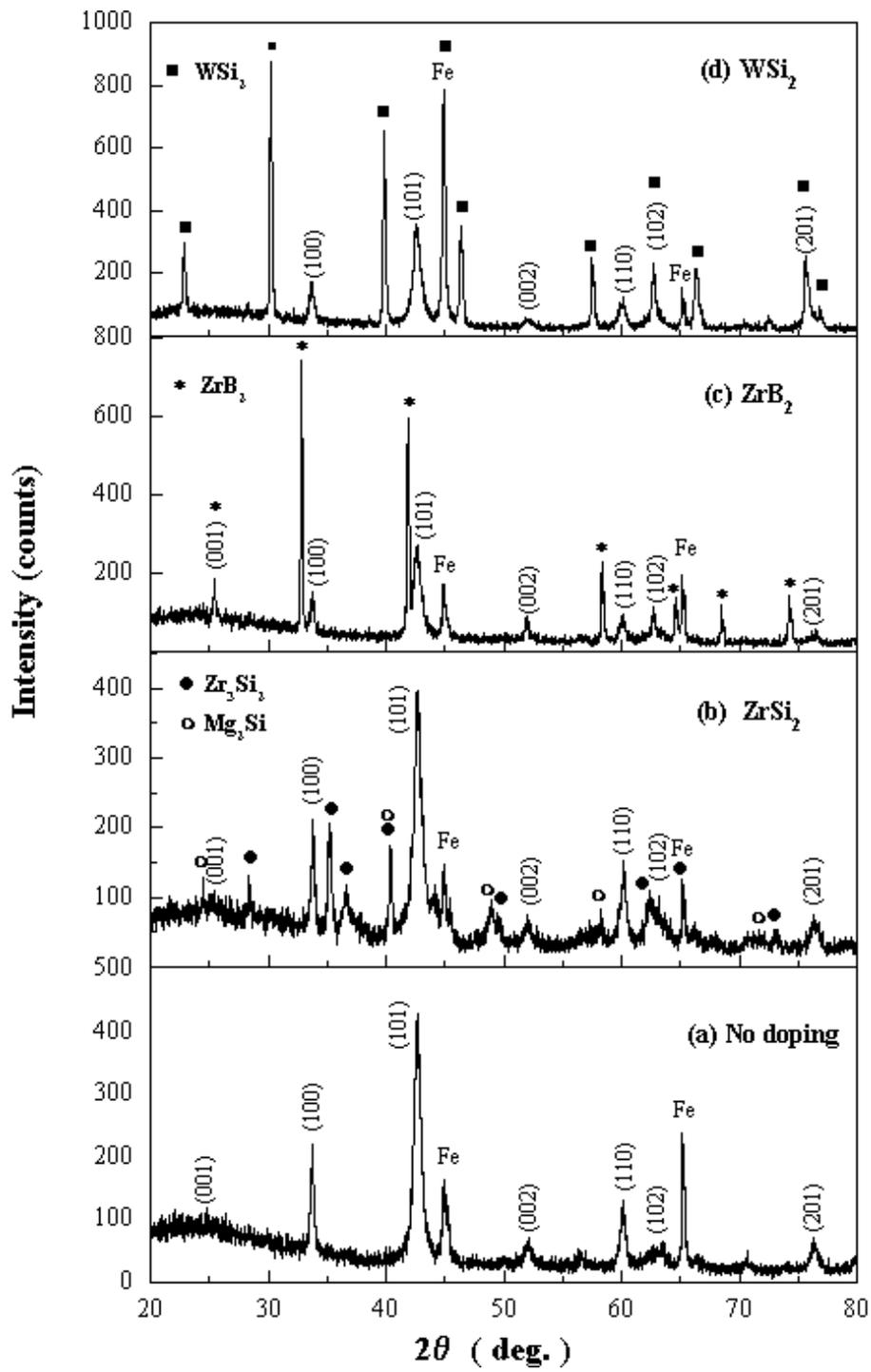

Fig. 1  Ma et al.

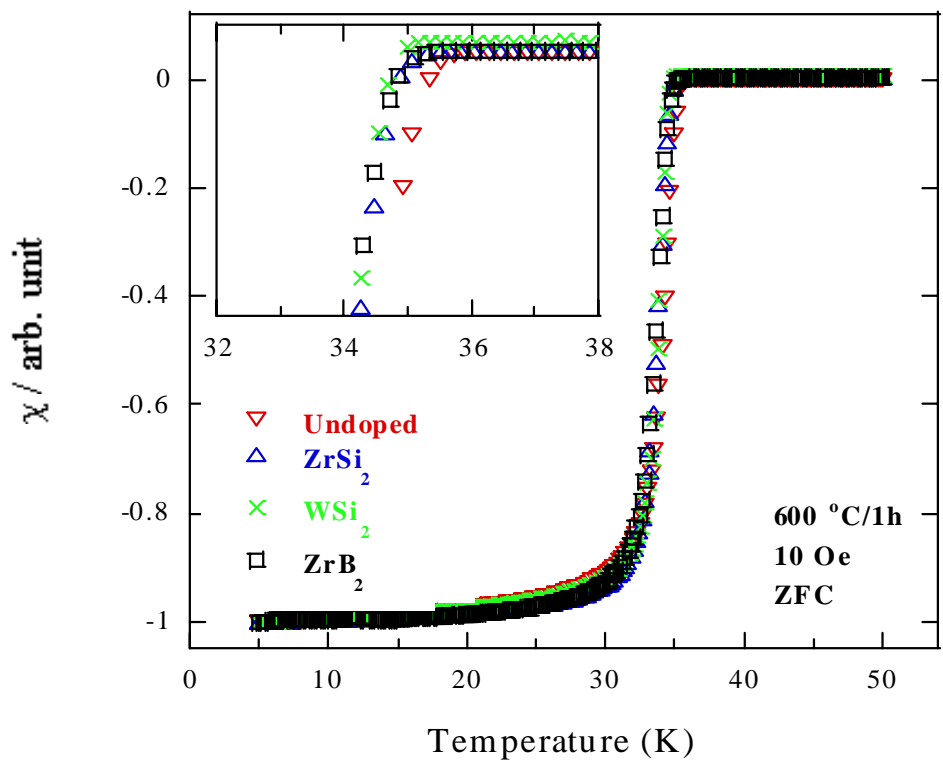

Fig. 2  Ma et al.

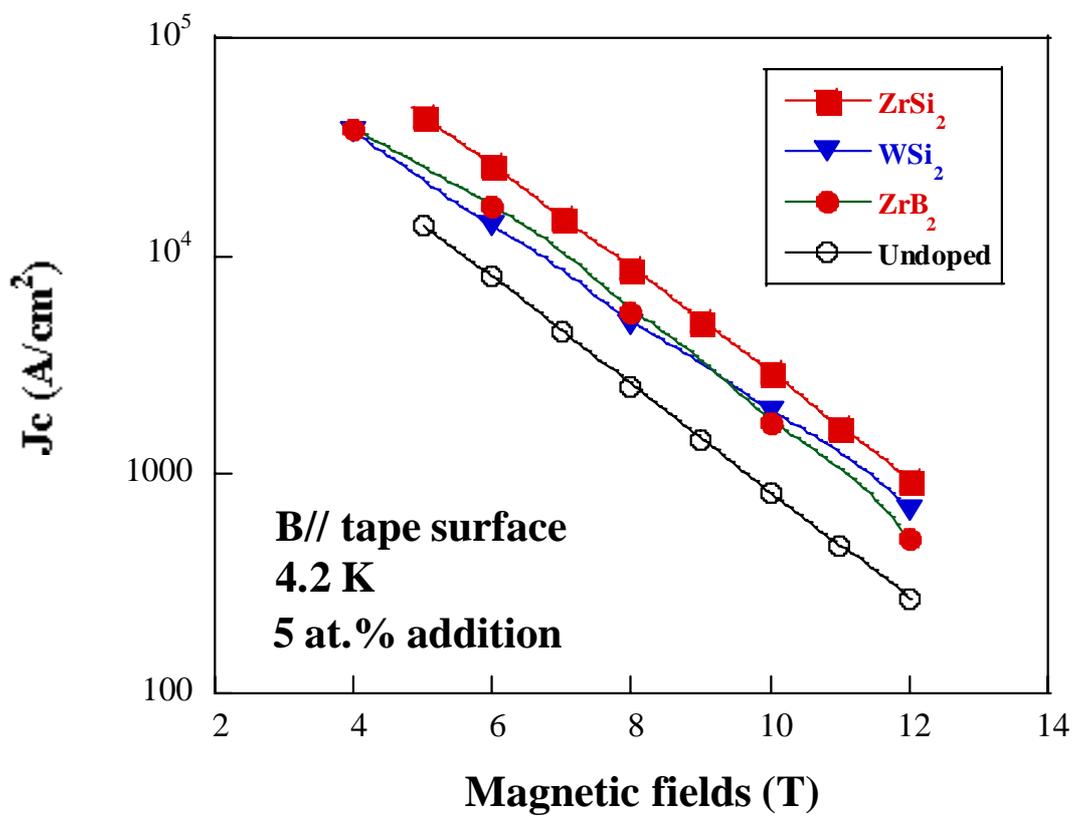

Fig. 3   Ma et al.

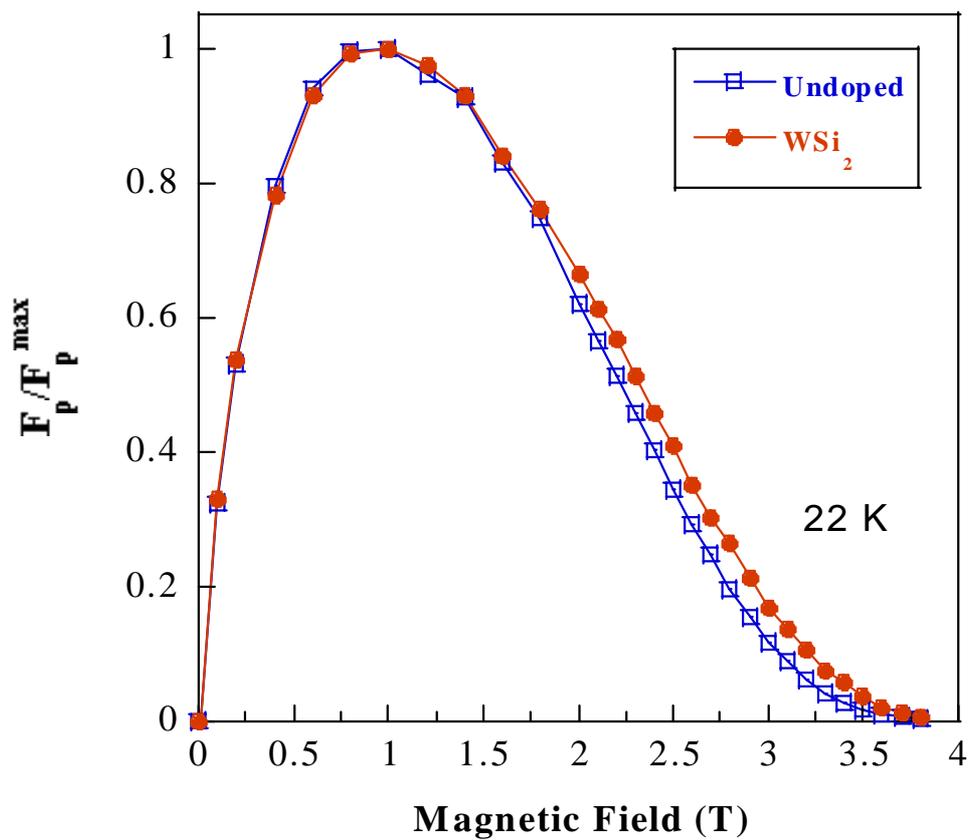

Fig. 4    Ma et al.

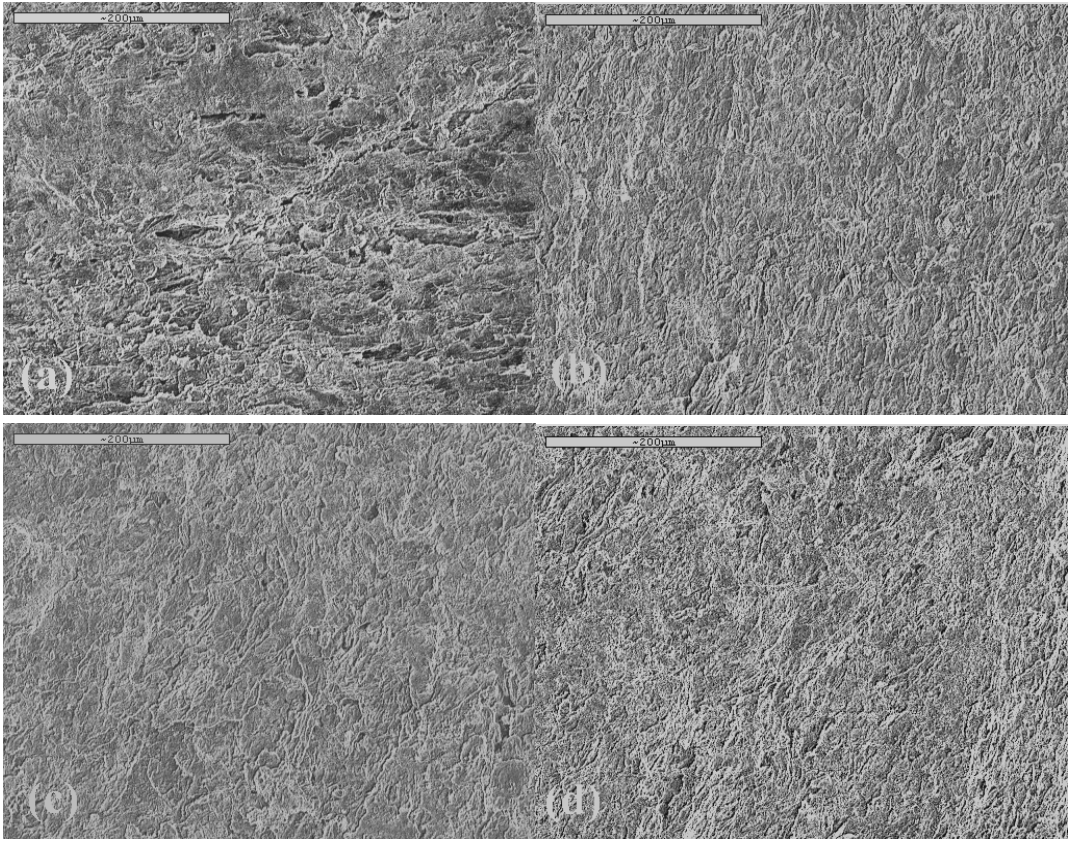

Fig. 5    Ma et al.